\documentclass[%
 reprint,
superscriptaddress,
 amsmath,amssymb,
 aps,
prl,
]{revtex4-2}

\usepackage{graphicx}
\usepackage{dcolumn}
\usepackage{bm}

\begin{document}


\title{Asymptotic mass limit of large fully-heavy compact multiquarks}

\author{M. C. Gordillo}
\email{cgorbar@upo.es}
\affiliation{Departamento de Sistemas F\'{\i}sicos, Qu\'{\i}micos y Naturales, Universidad Pablo de
Olavide, Carretera de Utrera km 1, E-41013 Sevilla, Spain}
\affiliation{Instituto Carlos I de F\'{\i}sica Te\' orica y Computacional, Universidad de Granada, E-18071 Granada, Spain.}

\author{J. M. Alcaraz-Pelegrina}
\email{fa1alpej@uco.es}
\affiliation{Departamento de F\'{\i}sica.
Universidad de C\'ordoba, Campus de Rabanales, Edif. C2 
E-14071 C\'ordoba, Spain}

\date{\today}

\begin{abstract}
The properties of fully-heavy  arrangements including a number of quarks between 5 and 12 were calculated 
within the framework of a constituent quark model by using a diffusion Monte Carlo technique.  
We considered only clusters in which all the quarks had the same mass,  and whose number of particles and antiparticles were
adequate to produce color singlets.  All the multiquarks were in their lowest possible values of $L^2$ and $S^2$ operators. 
This means that we considered only  color-spin wavefunctions that were antisymmetric with respect to the interchange
of {\em any} two quarks of the same type.  We found that in both all-$c$ and all-$b$ multiquarks,  the mass per particle levels off 
for arrangements with a number of quarks larger of equal than six.  The analysis of their structure implies that the fully-heavy multiquarks are compact structures.                
\end{abstract}

\maketitle


                 
Protons and neutrons are the basic constituents of atomic nuclei.  Quantum chromodynamics (QCD) is the theory that describe them as a composite set of quarks and gluons interacting through the strong force.  However, QCD is not limited to associations to the light quarks ($u,d$) that make up the nucleons, but extends to other types of particles collectively called hadrons.  Those hadrons can include or being totally made of heavier quarks ($s,c,b$).  

Unfortunately,  as of today, it is impossible to solve analytically the QCD equations and deduce the hadron spectrum.  Among the phenomenological QCD-inspired models designed to fill that gap the so-called quark model stands out. 
It considers only the valence quarks and antiquarks within the hadrons and was 
independently proposed by Murray Gell-Mann \cite{gellmann} and George Zweig \cite{zweig}.  Even though it was designed to account for the properties of mesons (one quark and one antiquark) and baryons (three quarks),  it also 
 opens the door to larger associations  of quarks such as tetra and pentaquarks \cite{hidden,open}.  We can even have hexaquarks, such as 
 the experimentally produced deuteron \cite{1932} and the well-established $d^*(2380)$ resonance \cite{d2380b,d2380c,d2380d,d2380e,d2380}.  The quark model does not impose,  in principle, any limit to the upper size of those clusters of quarks, and in this work we will use it to obtain the masses of all possible fully-heavy multiquarks.  Of all the possible compositions of those clusters, we will stick to arrangements in which all the quarks have the same mass. This means to consider (see below) sets up to 12 $c$ or $b$ quarks and/or antiquarks.  To do so,  we have to solve the Schr\"odinger equation derived from the Hamiltonian 
\cite{reviewvalcarce}: 
\begin{equation}
H = \sum_{i=1}^{N_q} \left( m_i+\frac{\vec{p}^{\,2}_i}{2m_i}\right) + \sum_{i<j}^{N_q} V(r_{ij}) \,,
\label{eq:Hamiltonian}
\end{equation}
where $N_q$ is the number of quarks,  while
$m_{i}$ and $\vec{p}_i$ are the mass and momentum of the $i$ quark. 
This a non-relativistic approximation, and it is expected to work best for the fully-heavy ensembles that will be considering in this work.  
To produce experimentally those multiquarks is, in principle, possible,  as the discovery of the X(6900) (though to be a fully $c$-tetraquark) attests \cite{x6900}. 
$V(r_{ij})$, is a two-body potential that depends only on the distance between quarks, $r_{ij}$, and can be written as the sum of one-gluon exchange term given by \cite{rujula:75,bhaduri:81} :  
\begin{equation}
V_{\text{OGE}}(r_{ij}) = \frac{1}{4} \alpha_{s} (\vec{\lambda}_{i}\cdot
\vec{\lambda}_{j}) [\frac{1}{r_{ij}} - \frac{2\pi}{3m_{i}m_{j}} \delta^{(3)}(r_{ij}) (\vec{\sigma}_{i}\cdot\vec{\sigma}_{j}) ] \,,
\end{equation}
that includes both Coulomb and hyperfine terms, and the lineal confining potential:
\begin{equation}
V_{\text{CON}}(\vec{r}_{ij}) = (b\, r_{ij} + \Delta) (\vec{\lambda}_{i}\cdot \vec{\lambda}_{j}). 
\end{equation}
that approximates the contribution of multigluon exchanges. $\vec{\lambda}$ and $\vec{\sigma}$ are the  Gell-Mann and Pauli matrices, respectively, 
and account for the color and spin degrees of freedom.  The Dirac delta function was regularized 
in the standard way  \cite{Semay:1994ht, SilvestreBrac:1996bg,prdyo1} in order to make possible the calculations.  The parameters needed to fully define the interaction 
were taken from Refs.  ~\cite{Semay:1994ht, SilvestreBrac:1996bg},  and were the same as the used in previous calculations 
for smaller clusters \cite{prdyo1,prdyo2,prdyo3,prdyo4}.  The masses of the hadrons computed with this potential were found to be in good agreement with 
experimental data, when available \cite{prdyo1}.  Since this non-relativistic approximation applies best to heavy quarks,  in order to describe light quarks (u, d or s) we would have to include additional terms \cite{vijande:05}, something that will not be done in this work. 


To solve the Schr\"odinger equation derived from the Hamiltonian in  Eq. \ref{eq:Hamiltonian}, we resorted to a  diffusion Monte Carlo (DMC) scheme \cite{kalos,boro94,hammond,spin-orbita,prdyo1}.  This will provide us with the desired masses of the ground states of the different set of quarks.  This method needs an initial approximation to the real many-body wavefunction of the clusters, the {\em trial function}, that should include  all the information known a priori about the different systems.  We chose the expression
\cite{prdyo1}:
\begin{eqnarray}
\Psi({\bf r_1, r_2},   \ldots,  {\bf  r_n}, s_1,s_2,  \ldots, s_n,c_1,c_2,\ldots, c_n) = \nonumber \\
\Phi ({\bf r_1, r_2},   \ldots,  {\bf  r_n}) \nonumber \\ 
\left[\chi_s (s_1,s_2,  \ldots, s_n) \bigotimes \chi_c (c_1,c_2,\ldots, c_n) \right] ,
\end{eqnarray}
where ${\bf r_i}$, $s_i$ and $c_i$ stand for the position, spin and color of the particle $i$, that is inside a cluster of $n$ quarks.
In this work, we are going to consider only multiquark states that are eigenvectors of the angular momentum operator, $L^2$ with 
eigenvalue $\ell$ = 0.  This means that $\Phi$ should depend on the distance between pairs of quarks and not on their absolute positions.  Following Ref. \onlinecite{prdyo1}, we have used:
\begin{equation} \label{radial}
\Phi ({\bf r_1, r_2},   \ldots,  {\bf  r_n}) = \prod_{i=1}^{N_q} \exp(-a_{ij} r_{ij}),
\end{equation} 
No other alternatives to the form of the radial part of the trial function were considered in this work since, in principle, the DMC algorithm should be able to correct its possible shortcomings and produce the exact masses of the arrangements \cite{hammond}. 
The $a_{ij}$ values were chosen in accordance to the boundary conditions of the problem \cite{prdyo1}.
 $\chi_s$ and $\chi_c$ are linear combinations of the eigenvectos of the spin and color operators defined by:
\begin{equation}
F^2 = \left(\sum_{i=1}^{N_q} \frac{\lambda_i}{2} \right)^2  
\end{equation}  
and 
\begin{equation}
S^2 = \left(\sum_{i=1}^{N_q} \frac{\sigma_i}{2}\right)^2.
\end{equation}
with eigenvalues $F^2$ = 0 (colorless functions) and $S$ = 0 or 1/2, depending on whether the number of quarks in the multiquark is ever or odd, respectively.  Those are the lowest possible eigenvalues for the spin operator and the only ones considered in this work. For instance, for the $cccc\bar{c}\bar{c}\bar{c}\bar{c}$ ($c^4 \bar{c}^4$) octaquark, we have 23 color and 14 spin functions meeting those criteria. This means 322 $\chi_s \bigotimes \chi_c$ possible combinations.  
That said,  we have to remember that since 
Eq. \ref{radial} is symmetric with respect to the exchange of any two {\rm identical} quarks, we have to produce spin-color combinations antisymmetric with respect to those exchanges, as befits to a set of fermions as quarks are.  
To do so, we apply the antisymmetry operator
\begin{equation}
\mathcal{A} = \frac{1}{N} \sum_{{\alpha}=1}^N (-1)^P \mathcal{P_{\alpha}}
\label{anti}
\end{equation}
to that color-spin set of functions.  Here, $N$ is the number of possible permutations of the set of quark indexes, $P$ is the order of the permutation, 
and $\mathcal{P_{\alpha}}$ represents the matrices that define those permutations.  
Once constructed the matrix derived from the operator in Eq. \ref{anti},  we have to check if we can find any eigenvector with eigenvalue equal to one.  If this is so,  those combinations will be the input of  the DMC calculation \cite{prdyo1}.  For the octaquark,  we have that of all the 322 color-spin functions, only 2 are antisymmetric with respect to the interchange of {\em all} the pairs of quarks and, separately,  of {\em all} the pairs of  antiquarks. 
The analysis of the eigenvectors of the antisymmetry operator indicates that there are no antisymmetric color-spin functions for structures in which any of the quark or antiquark subsets contains more than 6 units.  This means that the largest possible fully heavy multiquark is the $c^6 \bar{c}^6$ dodecaquark.  Moreover, neither the $c^9$ nonaquark nor the $c^7\bar{c}^4$ undecaquark, or their $b$-counterparts  are viable structures.  Independently, the $c^6\bar{c}^3$ nonaquark is also impossible since  no antisymmetric color-spin combinations  with respect to the interchange of any pair of $c$ quarks were found.


\begin{table} 
\caption{\label{tab:table1} Masses of the mutiquarks considered in this work in MeV.  The error bars are given in parenthesis. 
}
\begin{tabular}{cccc} \hline 
          &  $c^4\bar{c}$  & $c^3\bar{c}^3$  & $c^5\bar{c}^2$  \\ 
Mass & 8195(2) &  9614(2) & 11543(4)   \\   
           & $c^4\bar{c}^4$ & $c^5\bar{c}^5$ & $c^6\bar{c}^6$      \\  
Mass &   13133(4) & 16539(4) &  19808(4) \\ \hline \hline
            &  $b^4\bar{b}$  & $b^3\bar{b}^3$  & $b^5\bar{b}^2$  \\ 
 Mass  &  24211(2) &  28822(2) & 33970(4) \\
            & $b^4\bar{b}^4$ & $b^5\bar{b}^5$ & $b^6\bar{b}^6$      \\ 
  Mass          & 38815(4) &  48599(4) & 58232(4) \\ \hline
\end{tabular} 
\label{table1}
\end{table}

The masses of the multiquarks obtained by the DMC method are given in Table \ref{table1}.  As indicated above,  all are colorless clusters with $S$=0 or 1/2 depending on whether the total number of quarks is even or odd,  respectively.  We have to stress that the color-spin functions used in the calculations are the eigenvalues of the antisymmetry operator given in Eq. \ref{anti}, with no quark groupings different to those that put together identical particles.  For instance,  in pentaquarks, we do not consider baryon+meson or diquark+diquark+antiquark arrangements \cite{jorge,penta3},  but a function that is antisymmetric with respect to the exchange of {\em any} pair of the four quarks considered to be undistinguishable.   In any case, the results for that particular multiquark are virtually identical to those of Ref. \onlinecite{penta1}, in which the same function is used. Those results validate our approach,  that 
allows us to dispense with Young-tableaux diagrams to calculate larger clusters. 

\begin{figure}
\begin{center}
\includegraphics[width=0.8\linewidth]{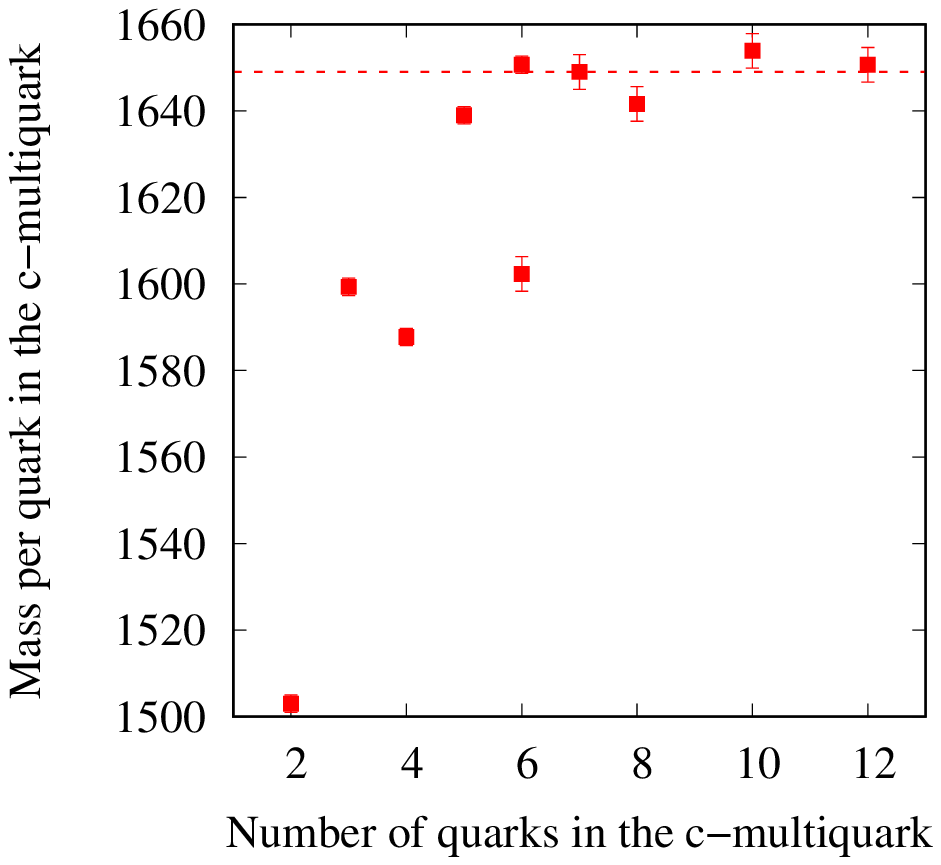}
\caption{Mass per particle in all the $c$-multiquarks considered in this work.  
Where not shown,  error bars are of the size of the symbols.  The data for the meson,  baryon and tetraquark are taken from Ref. \onlinecite{prdyo1}, while the upper symbol for the hexaquark corresponds to the mass of the open charm hexaquark given in Ref. 
\onlinecite{prdyo4}.  The dashed line represents the average mass for the open charm hexaquark , and the hepta-, octa-, nona-, deca- and docedaquarks. }
\label{fig1}
\end{center}
\end{figure}

\begin{figure}
\begin{center}
\includegraphics[width=0.8\linewidth]{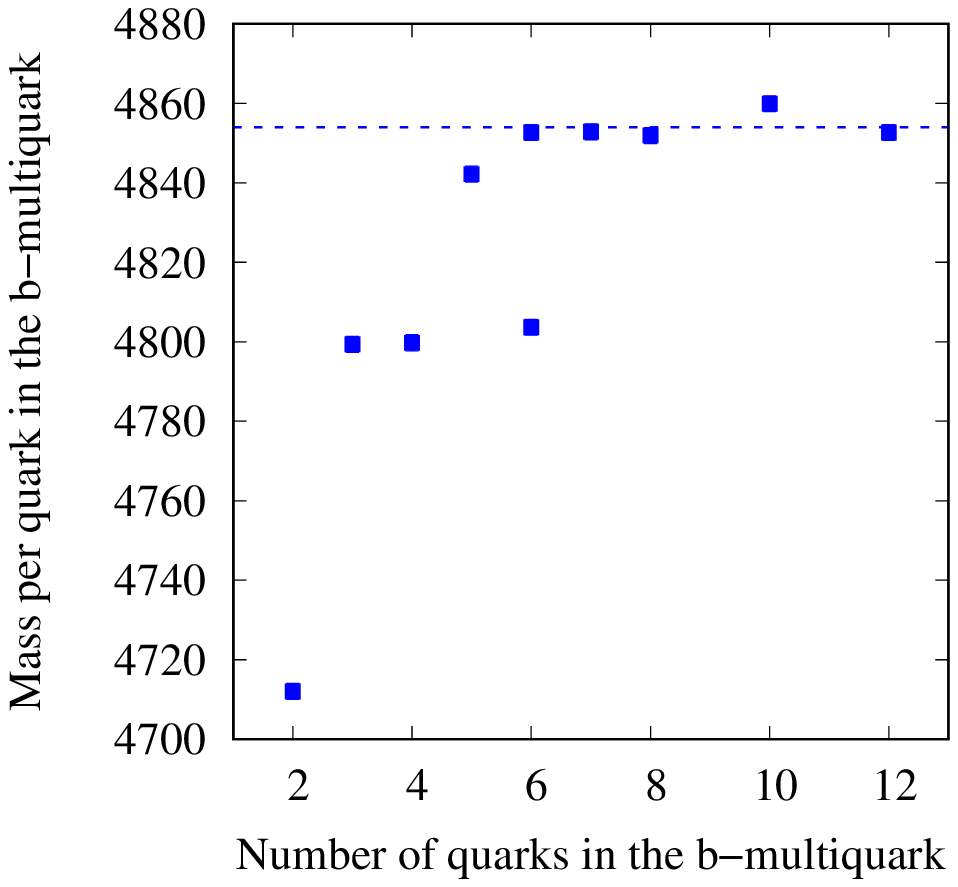}
\caption{Same as in the previous figure,  but for $b$-multiquarks. Error bars are of the size of the symbols and not shown for simplicity. The dashed line have the same meaning as in Fig. \ref{fig1} but for the $b$-multiquarks.  The source of the data is the same as for their $c$-counterparts. 
}
\label{fig2}
\end{center}
\end{figure}

To better visualize the results in Table \ref{table1},  we display the mass per particle as a function of the number of particles in the cluster in Figs. \ref{fig1} and \ref{fig2}.  The data not given in Table \ref{table1} are taken from Refs. \onlinecite{prdyo1} and \onlinecite{prdyo2}. 
Something is immediately apparent: from the open-charm hexaquark up, the mass per particle of the clusters reaches a plateau both for $c$- and $b$-multiquarks.  This basically means that to modify the number of quarks beyond six, we will have to increase the mass of the system by a constant value of 1649 and 4854 MeV per particle for $c$- and $b$-multiquarks, respectively. 

\begin{figure}
\begin{center}
\includegraphics[width=0.8\linewidth]{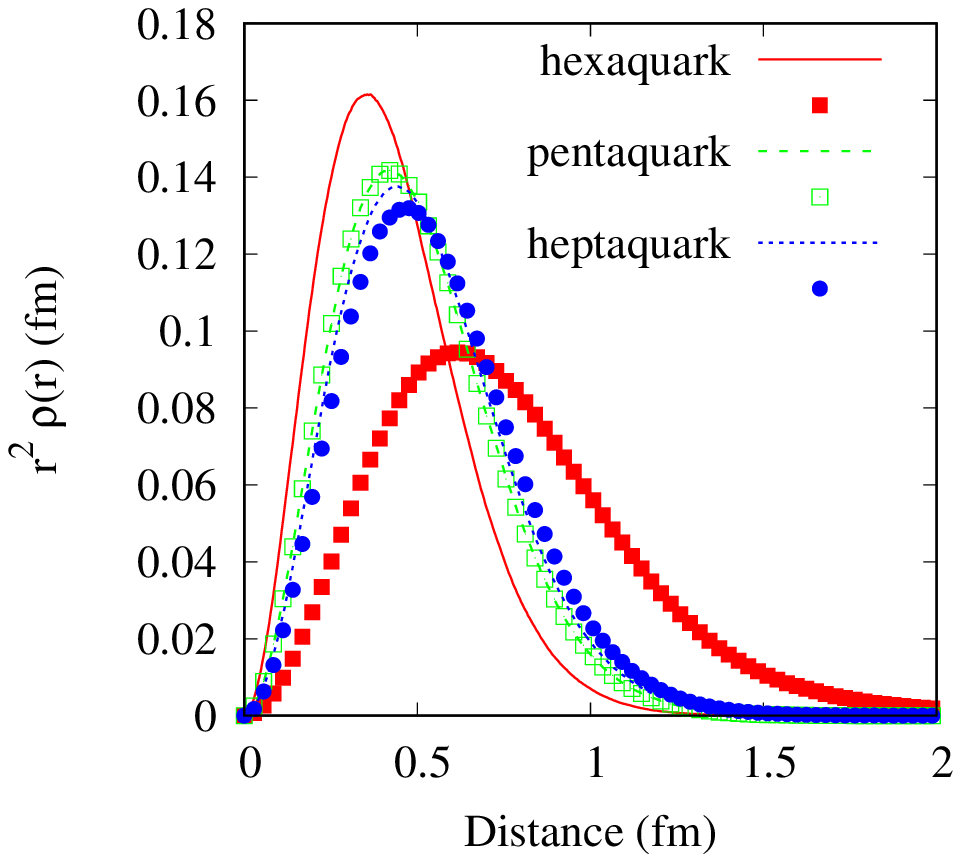}
\caption{Radial distribution functions for different particle pairs in $c$-multiquarks. Solid lines,  $c-c$ pairs (equivalent to $\bar{c}-\bar{c}$);  symbols,  $c-\bar{c}$ pairs. 
The hexaquak displayed correspond to the $c^3\bar{c}^3$ hidden charm structure.
All distributions are normalized to one.}
\label{fig3}
\end{center}
\end{figure}

\begin{figure}
\begin{center}
\includegraphics[width=0.8\linewidth]{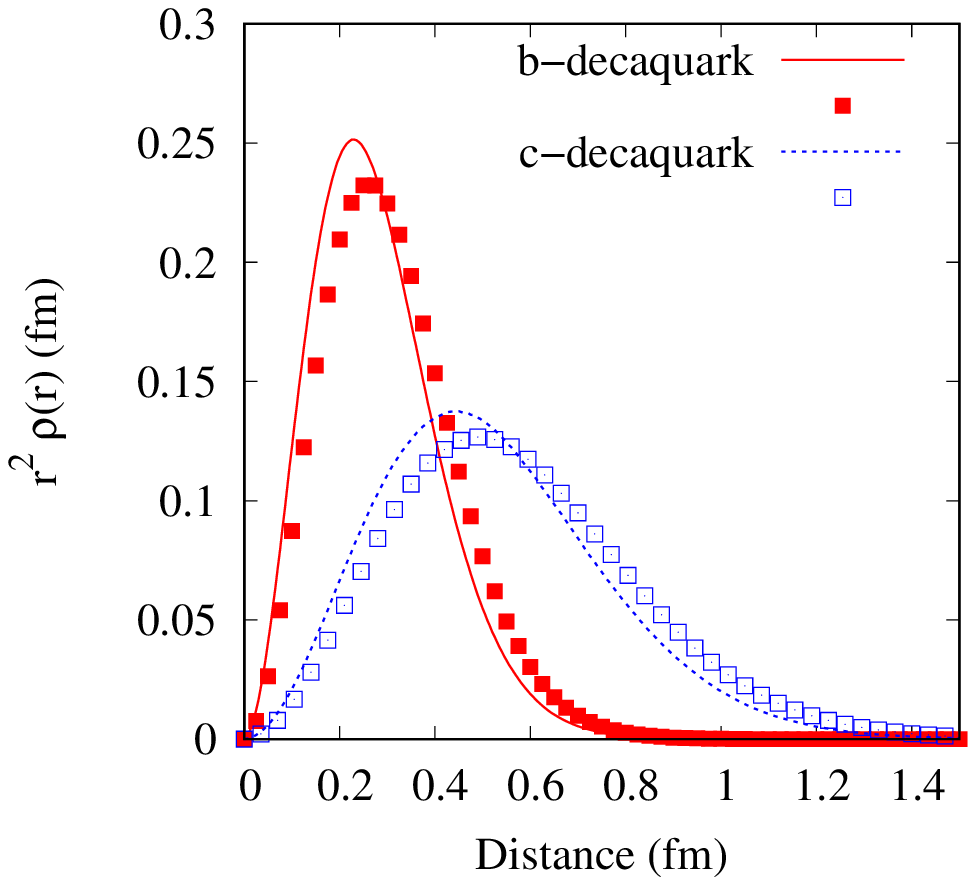}
\caption{Comparison between the pair distribution functions for a $c$-decaquark and a $b$-decaquark.  Lines,  $q-q$ pairs;  symbols,  $q-\bar{q}$ pairs. 
}
\label{fig4}
\end{center}
\end{figure}

The structure of the clusters can be deduced from the radial distribution functions, depicted in Figs. \ref{fig3} and \ref{fig4}.  Those give us the probabilities of having another particle at a particular distance of a given one.  We show only the more representative structures, the remaining ones being similar to those displayed.  First, we can see that all the clusters are compact structures, i.e., the probability of finding another particle at distances beyond a maximum of 2 fm goes rapidly to zero.  In addition,  in the majority of cases there is very little difference between the probability of finding another quark (solid lines) or an antiquark (symbols) for any particule at a given distance.   This is similar to what happens for smaller multiquarks \cite{prdyo1,prdyo2}. The only exception is the hidden-charm hexaquark, in which the $c-c$ and $c-\bar{c}$ are noticiably different, and in which the first of them is virtually identical to the corresponding to the $ccc$ baryon.  The reason is that in that system,  the quarks and antiquarks group to produce a baryon and an antibaryon glued together. The same happens with the $b^3\bar{b}^3$ system. 

In this work we have 
calculated the color-spin functions with an algorithm that dispense with the need to use of  Clebsch-Gordan coefficients.  This is necessary since the increase in the number of color-spin functions with the number of quarks makes that approximation impossible.  For instance, for an heptaquark, we have 11 color and 14 spin functions that make a total of 154 combinations.  This is to be compared with the 15 color-spin possibilities for a pentaquark \cite{penta1} or the 25 for an open-charm hexaquark \cite{5f0,hexa2}.  The use of this technique in combination with a DMC algorithm, originally developed to deal with many-body systems, allowed us obtain the masses of all possible fully-heavy s-wave multiquarks.   What we found is that,  from a number of quarks beyond six, the mass of those systems is linearly proportional to the number of particles in the arrangements, i.e.,  in relative terms, there is no mass penalty in producing progressively larger multiquarks, as it is in going from a meson to a tetraquark.  This means that, in mass terms, is equally probable to have an open-charm hexaquark as to produce an heptaquark or octoquark.  

\begin{acknowledgments}

We acknowledge financial support from Ministerio de
Ciencia e Innovación MCIN/AEI/10.13039/501100011033 
(Spain) under Grant No. PID2020-113565GB-C22
and from Junta de Andaluc\'{\i}a group PAIDI-205.  
We also acknowledge the use of the C3UPO computer facilities at the Universidad
Pablo de Olavide.
\end{acknowledgments}

\bibliography{multiquarks2}

\end{document}